\documentclass[aps,prl,twocolumn,superscriptaddress,showpacs]{revtex4}

\usepackage{mathrsfs}
\usepackage{graphicx}
\usepackage{color}
\usepackage{amsmath}
\usepackage{amssymb}

\begin{document}

\title{Collective dynamics effect transient subdiffusion of inert tracers in gel
networks}

\author{Alja\v{z} Godec}
\email{aljazg@agnld.uni-potsdam.de}
\affiliation{Institute of Physics \& Astronomy, University of Potsdam, 14776
Potsdam-Golm, Germany}
\affiliation{National Institute of Chemistry, 1000 Ljubljana, Slovenia}
\author{Maximilian Bauer}
\affiliation{Institute of Physics \& Astronomy, University of Potsdam, 14776
Potsdam-Golm, Germany}
\affiliation{Physics Department, Technical University of Munich, Garching, Germany}
\author{Ralf Metzler}
\email{rmetzler@uni-potsdam.de}
\affiliation{Institute of Physics \& Astronomy, University of Potsdam, 14776
Potsdam-Golm, Germany}
\affiliation{Department of Physics, Tampere University of Technology, FI-33101
Tampere, Finland}

\date{\today}

\begin{abstract}
Based on extensive Brownian dynamics simulations we study the thermally driven
motion of a tracer bead in a cross-linked, dynamic gel network in the limit when
the tracer bead's size is of the same size or even larger than the equilibrium
mesh size of the gel. The analysis of long individual trajectories of the tracer
bead demonstrates the existence of pronounced transient anomalous diffusion,
accompanied by a drastic slow-down of the gel-bead relaxation dynamics. From the
time averaged mean squared displacement and the van Hove cross-correlation
function we elucidate the many-body origin of the non-Brownian tracer bead
dynamics. Our results shed new light on the ongoing debate over the physical
origin of sterical tracer interactions with structured environments.
\end{abstract}


\maketitle

Modern single particle tracking technology \cite{SPT} unveils the non-Brownian
stochastic motion of submicron tracers or fluorescently labeled macromolecules
in a variety of complex liquids. Anomalous diffusion of the form $\langle\mathbf{r}
^2(t)\rangle\simeq t^{\alpha}$ with $0<\alpha<1$ \cite{report} was observed in the
cytoplasm and membrane \cite{mrna,granules,membrane} of living biological cells,
as well as in dense polymer or protein solutions \cite{solutions}. Causes for the
observed subdiffusion may be the crowded state of the environment \cite{crowding}
or sticking effects between the tracer and the environment \cite{xu}. Another
important origin for anomalous diffusion is the subtle interaction between the
tracer particle and the dynamic confines of a structured, gel-like environment
with a well defined mesh size \cite{wong}.

The existence of a mesh-like, structured environment is a defining property for
the transport in several systems. Thus, biological cells are internally equipped
with a characteristic mechanical network consisting of actin and other biofilaments
through which submicron particles diffuse or are actively transported
\cite{philips}. Inside the eukaryotic nucleus, biomolecules diffuse through the
complex chromatin network \cite{chromatin}. In cellular tissues the space between
cells is filled with the mesh-like extracellular matrix \cite{ecm}. Novel mobile
clinical diagnosis tools are based on hydrogel films, into which pathogens such as
viral particles need to diffuse \cite{ttl}. Finally, particles in a biofilm move
through a flexible, porous bacterial matrix \cite{biofilm}.

\begin{figure}
\includegraphics[width=6.cm]{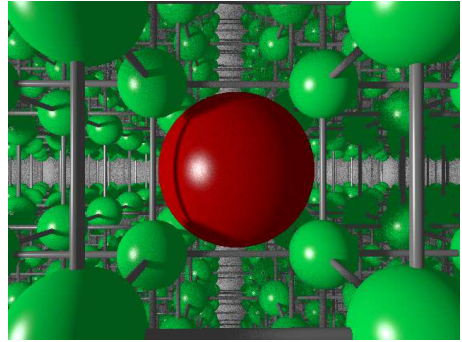}
\caption{Schematic of the tracer particle (red), in a network of gel beads (green)
connected by Morse springs (gray sticks).}
\label{schm}
\end{figure}

How do particles diffuse through a flexible, thermally agitated mesh such as a
hydrogel? The answer is trivial as long as the diffusing particles are either
much smaller or much larger than the typical mesh size. In these cases they
diffuse normally or are completely immobilized in the mesh \cite{wong}. The
physically intriguing case is met when the size of the diffusing particles is
comparable to the typical mesh size. This is the scenario we consider here
(Fig.~\ref{schm}). From extensive Brownian dynamics simulations we
elucidate the fundamental micro- and mesoscopic physical principles behind the
(transient) anomalous tracer dynamics on the collective many-body level. The
tracer's mean squared displacement (MSD) and the van Hove cross-correlation
function demonstrate the occurrence of significant tracer subdiffusion and
massive cooperative breathing effects of the mesh, allowing relatively large
particles to move in the gel, albeit under massively reduced diffusive progress.

\emph{Tracer motion in gels.} The motion of a tracer in a flexible mesh can be
viewed as a convolution of the Brownian self-dynamics of the tracer and the
thermal agitation of the interacting gel particles, important ingredients
characteristic for the tracer-gel system being the sterical obstruction of the
tracer by the mesh \cite{ster}, tracer-gel interactions \cite{inter}, and the
elastic response of the network itself \cite{elastic}. Remarkably, distinct
anomalous tracer diffusion was observed in numerous experiments \cite{wong,Rouse,
DLS,inter,BPJ} and simulations \cite{sims}. To explain this non-Brownian dynamics
in the gel two scenarios are typically invoked: (i) In the first scenario trapping
of the tracer occurs due to attractive tracer-gel interactions, that effect
 transient binding to the gel network \cite{Rouse,inter,Saxton,Holm}, such that
the tracer intermittently follows the gel chains' Rouse  dynamics \cite{Rouse} or,
alternatively, remains transiently immobilized for a given period \cite{Saxton}.
(ii) In the second scenario sterical hindrance within a fractal organization of
the accessible volume is suggested to cause the non-Brownian tracer dynamics
\cite{wong,DLS,inter,BPJ,sims,Lang}. In these approaches the gel is represented by
impenetrable beads (either static \cite{sims} or dynamic \cite{Lang}), positioned
on a lattice. For static networks simulations revealed that isolated, randomly
positioned beads give rise to non-transient subdiffusion, whereas tracer diffusion
in presence of locally clustered beads in the form of condensed cubes randomly
distributed in space is only transiently anomalous \cite{sims}. The debate on the
dominant mechanism in real gel systems remains open \cite{inter}. Notably, both
mechanisms are only explained on a heuristic level and a better physical
understanding is needed.

While lattice representations of gel structures offer a computationally convenient
tool to study tracer diffusion therein \cite{sims,Lang}, they suffer from the
severe unphysical localization of the gel despite its inherent flexibility and
exposure to a thermostat. Moreover, a static gel introduces spurious artifacts
in the tracer dynamics, as quenched configurational disorder of obstacles itself
introduces a strongly persistent memory in the motion of a Brownian particle at
all obstacle densities \cite{Frey}. As we will show here, the tracer motion is
inherently related to the breathing of the gel, i.e., collective soft modes of
local expansion and contraction of the thermalized mesh. Such breathing modes
cannot be appropriately captured in the lattice models. We thus face a pressing
need for more realistic continuum models with fully coupled, many-body tracer-gel
dynamics to gain physical insight into the effective tracer motion in such systems.

\emph{Model.} We consider a gel of $7\times7\times7$ connected beads and a freely
diffusing tracer bead (Fig.~\ref{schm}) interacting  with the repulsive part of
the shifted Lennard-Jones potential
\begin{equation}
\label{LJ}
V^{ij}_{LJ}(r_{ij})=4\epsilon_{ij}\left[\left(\frac{\sigma_{ij}}{r_{ij}}\right)
^{12}-\left(\frac{\sigma_{ij}}{r_{ij}}\right)^6\right]\Theta(2^{1/6}\sigma_{ij}
-r_{ij})+\epsilon_{ij},
\end{equation}
where $i$ and $j$ for a specific pairwise interaction take on the values $t$ for
the tracer particle and $g$ for gel beads. $r_{ij}\equiv|\mathbf{r}_i-\mathbf{r}
_j|$ is the physical separation between $i$ and $j$. $\sigma_{ij}$ are the
separations at zero unshifted potential, i.e., $\sigma_{ii}$ is the diameter of
species $i$. $\epsilon$ sets the respective energy scales, and $\Theta(x)$ is the
step function. Nearest neighbor gel beads are connected by Morse springs with the
potential
\begin{equation}
\label{Morse}
V^{ij}_{M}(r_{ij})=\varphi_d\left(e^{-2\zeta d^e_{ij}}-2e^{-\zeta d^e_{ij}}\right)
\Theta(r_c-r_{ij})-V_s,
\end{equation} 
with $\zeta=\sqrt{k_e/(2\varphi_d)}$. $\varphi_d$ is the potential depth, $k_e$
the force constant at the potential minimum, and $d^e\equiv r_{ij}-r_{eq}$,
where $r_{eq}$ denotes the equilibrium separation. $r_c$ is a cutoff distance
and $V_s=V^{ij}_{M}(r_c)$. The particle positions are governed by the
overdamped Langevin equations
\begin{equation}
\label{Langevin}
\frac{d\mathbf{r}_i(t)}{dt}= \beta D_i \sum_{j\ne i} \mathbf{F}_{ji} +
\sqrt{2D_i}\boldsymbol{\xi}(t),
\end{equation} 
where $D_i$ is the diffusion coefficient of the tracer or gel bead, $\beta^
{-1}=k_BT$, and $\mathbf{F}_{ji}=-\nabla_{\mathbf{r}_i}U_{ij}(r_{ij})$ is the
force acting on particle $i$ due to particle $j$, and equals the sum of the
Lennard-Jones potential (\ref{LJ}) and the Morse potential (\ref{Morse}).
$\boldsymbol{\xi}(t)$ is a delta-correlated Gaussian noise with
zero mean and component-wise variance $\langle\xi_k(t)\xi_l(t')\rangle=\delta(t-
t')\delta_{k,l}$. Since hydrodynamic interactions are screened in dense systems
\cite{Adelman} we neglect them in our model. We use dimensionless units and
express distances in units of the gel bead diameter $\sigma_{gg}$, energies in
units of $\beta^{-1}$ and diffusivities in units of $\sigma_{gg}/\sqrt{\epsilon
_{gg}}$, thus fixing the time unit $t_1=1$. We take $\epsilon_{gg}=1.2$
corresponding to good solvent conditions for the corresponding linear chains of
beads \cite{Zhou}, and we set $\epsilon_{tg}=\epsilon_{gg}$. We choose $r_{eq}=4$,
which also fixes the gel unit cell size $L_0=r_{eq}$. Finally, we consider various
sizes $\sigma_{tt}$ of the tracer particle and spring stiffness $k_e$. The free
tracer diffusivity is given by the Stokes-Einstein relation, $D\simeq1/\sigma_{tt}$.
We use combining rules such that $\sigma_{tg}=\frac{1}{2}(\sigma_{gg}+\sigma_{tt}
)$. The Langevin equations are solved with the Euler method with integration step
$10^{-4}$ under periodic boundary conditions. The initial configuration is chosen
randomly to be slightly displaced from the energetically minimal configuration,
not allowing significant overlap nor significant bond stretching. After extensive
equilibration time of $10^2$, trajectories of length $10^4$ are simulated and
analyzed. In the following we drop the indices and use the remaining symbols for
the tracer particle $\sigma_{tt}\equiv\sigma$ etc. The overall time to simulate
and analyze a single trajectory of length $T=10^4$ was 5-6 weeks of single-CPU
time on a computer cluster, and our study thus reaches the current computational
limit.

\emph{Analysis  of  tracer  dynamics.} We use the time averaged MSD at lag time
$\Delta$ over a trajectory of length $T$,
\begin{equation}
\label{TAMSD}
\overline{\delta^2(\Delta)}=\frac{1}{T-\Delta}\int_0^{T-\Delta}\left[\mathbf{r}
(t+\Delta)-\mathbf{r}(t)\right]^2dt,
\end{equation}   
to characterize the tracer dynamics. For a wide range of parameters we observe
that $\overline{\delta^2}$ exhibits a distinct, transient subdiffusion between
the initial free diffusion regime and the asymptotically normal diffusion
regime at long lag times, see below. This behavior agrees with similar simulations
\cite{Zhou}. Being mainly interested in exploring the generic qualitative
features of the transient subdiffusion regime, which turn out to be essentially
independent of the chosen parameter values, in the following we focus on the
particular value $k_e=5$ characteristic for a moderately soft gel. The value of
$k_e$ only affects the values of the short and long time diffusivities and the
time window of the intermediate-asymptotic subdiffusion regime.

\begin{figure}
\includegraphics[width=6.5cm]{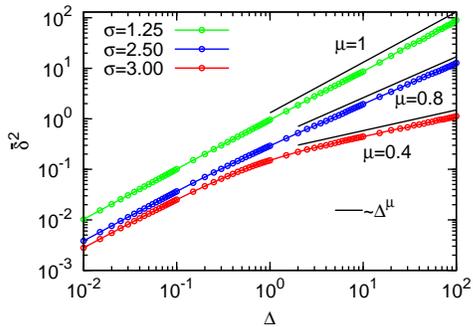}
\caption{Time averaged MSD of tracers of various radii $\sigma$ as function of
lag time $\Delta$. Full lines denote the intermediate-asymptotic power-law
scaling $\overline{\delta^2}\simeq\Delta^{\mu}$. Each set of results represents
an average of $\overline{\delta^2}$ over 10 different realizations.}
\label{tamsd}
\end{figure}

The typical behavior of the MSD $\overline{\delta^2}$ is shown in Fig.~\ref{tamsd}.
Since our model gel structure is comparatively open ($r_{eq}=4$), in agreement with
intuition tracer particles of size comparable to the gel bead diameter almost do
not feel the presence of the gel, and the transition from short to long time free
diffusion is barely visible on the double-logarithmic scale. In contrast, as the
size of the tracer grows, a pronounced subdiffusive power-law scaling $\overline{
\delta^2}\simeq \Delta^{\mu}$ emerges for $\Delta>1$, and the anomalous diffusion
exponent $\mu$ delicately depends on the tracer size. We note that the displacement
amplitude $\chi=|\mathbf{r}(t+\Delta)-\mathbf{r}(t)|$ on this time scale is several
times larger than the unit cell of the gel, and we hence observe a significant
tracer motion and not simply fluctuations around a localization. The latter will
be shown explicitly below. After the transiently anomalous regime there follows
the normally diffusive long time regime (not shown) \cite{comm}. These results
demonstrate that the existence of spatial obstructions alone is indeed sufficient
to effect transient anomalous diffusion, but they do not provide any details about
the underlying dynamic mechanisms.

\begin{figure}
\includegraphics[width=8.8cm]{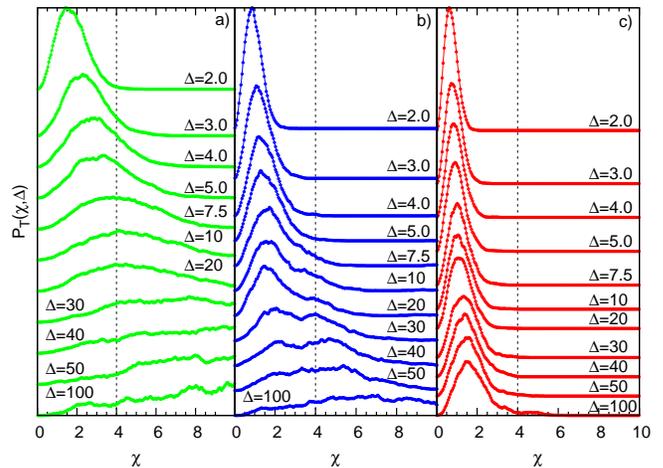}
\caption{Distribution $P_T$ for displacement amplitudes $\chi$ at lag time
$\Delta$, Eq.~(\ref{rdisp}), for three different tracer sizes: a) $\sigma=1.25$
(green), b) $\sigma=2.5$ (blue), and c) $\sigma=3$ (red). The dashed vertical
line in each panel denotes the approximate size of the unit cell of the gel mesh.
For visual convenience the curves are shifted vertically. Note that the scales
increase from a) to c), as each curve is normalized. Each set of results is from
an average over 10 different realizations.}
\label{rdpdf}
\end{figure}

To gain insight beyond the $\Delta$-scaling of the MSD $\overline{\delta^2}$ we
compute the distribution of relative displacements of magnitude
$\chi$ for given lag time $\Delta$ along the trajectory,
\begin{equation}
\label{rdisp}
P_T(\chi,\Delta)=\frac{1}{T-\Delta}\int_0^{T-\Delta}\delta(|\mathbf{r}(t+\Delta)-
\mathbf{r}(t)|-\chi)dt.
\end{equation} 
The results for this quantity for different tracer sizes are shown in
Fig.~\ref{rdpdf}. As expected from the behavior of $\overline{\delta^2}$ small
tracer particles do not feel appreciable obstruction by the gel mesh, their
relative displacement $\chi$ evolves smoothly as a single peak, which moves to
progressively higher values while simultaneously broadening (Fig.~\ref{rdpdf}a)).
With growing tracer size, however, the major fraction of relative displacements
represented by the dominant peak of $P_T$ spreads much more slowly and remains
almost localized even for longer $\Delta$, in particular, for $\sigma=3$, as
seen in Fig.~\ref{rdpdf}b) and c). Moreover, $P_T$ slowly develops a second,
larger displacement fraction at $\chi>3.5$ for $\sigma=2.5$ and $\chi>3$ for
$\sigma=3$, respectively. We stress that the tracer is not localized in the same
mesh cell on its trajectory, not even for the case $\sigma=3$, but ventures into
vicinal cells. We also point out that there exist an apparent time scale when
most escape attempts across the mesh boundary to the next mesh cell become
successful, corresponding to the spreading of the initial peak beyond the
confines of a mesh cell. This time scale is of the order of $\Delta\approx50$ for
tracers of size $\sigma=2.5$ and significantly longer than $\Delta\approx100$ for
$\sigma=3$, suggesting that it is the very cell escape dynamics that bestows the
subdiffusive nature on the tracer dynamics. A finite characteristic escape time
restores the terminal normal diffusion.

\emph{Collective gel-tracer dynamics.} Are the tracer passages from one mesh cell
to the next simply governed by an averaged obstruction effect and thus decoupled
from the breathing dynamics of the gel, or do they correspond to a much richer
scenario involving many-body effects? We study this question via the time
averaged van Hove cross-correlation function (vHCF)
\begin{equation}
\label{Hove}
G_T(r,\Delta)=\frac{\mathcal{C}^{-1}}{T-\Delta}\int_0^{T-\Delta}\sum_{i=1}^{N_{
gel}}\delta(|\mathbf{r}^b_i(t+\Delta)-\mathbf{r}(t)|-r)dt.
\end{equation} 
Here $\mathbf{r}^b_i(t)$ is the position of the gel bead $i$ at time $t$ and
$N_{gel}$ the overall number of gel beads. $\mathcal{C}^{-1}$ was chosen such
that the probability density of observing a gel bead infinitely far away from
the initial position of the tracer is unity at any lag time and corresponds to
dividing the uncorrected $G_T(r,\Delta)$ by the number density of beads, $\rho=
N_{gel}/V$. $G_T(r,0)$ corresponds to the static pair correlation function.
The vHCF effectively measures the memory loss of the environment about the
instantaneous tracer location and thus reflects the correlations of the
dynamics of the gel beads and the tracer particle.

\begin{figure}
\includegraphics[width=8.8cm]{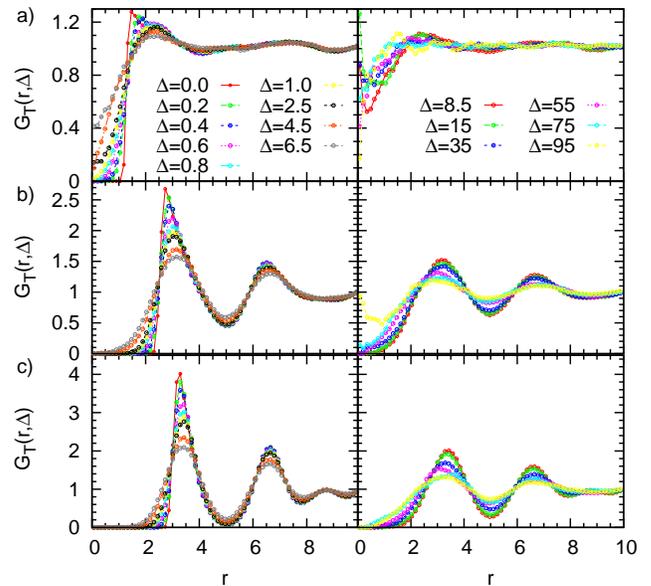}
\caption{vHCF for different tracer sizes at various lag times: a) $\sigma=1.25$,
b) $\sigma=2.5$, and c) $\sigma=3$. For visual convenience the curves were
divided in two panels according to $\Delta$.}
\label{Vhove}
\end{figure}

The results for $G_T(r,\Delta)$ for different tracer sizes are shown in
Fig.~\ref{Vhove}. In part a) for the smallest tracer, gel beads quickly
penetrate the tracer's initial position on the scale of $\Delta\sim1$,
during which the tracer typically moved by $\chi\sim1$, just about a bead
diameter (compare Figs.~\ref{tamsd} and \ref{Vhove}a)). The sharper peak
corresponding to nearest gel beads quickly decorrelates. On time-scales
$\Delta>10$ the small tracer typically moves a distance of (several) unit
cells, and $G_T(0,\Delta\gtrsim10)$ shows oscillations corresponding to vivid
fluctuations of gel bead positions. On these and longer time scales the
correlations are completely lost at distances beyond a very short cutoff,
which is due to the fact that the gel motion is translationally confined by
the elastic interactions. Hence, beyond a typical time scale of $\Delta\gtrsim10$
the  beads completely forget that the tracer particle was in their proximity.

The situation changes dramatically for larger beads. For $\sigma=2.5$ we observe
in Fig.~\ref{Vhove}b) that the position correlations are quite persistent, and
even the position of second nearest neighbor beads is still appreciably correlated.
We note that the tracer diameter easily fits a unit mesh cell. Even on a time 
scale of $\Delta\sim 10$, on which the tracer has likely already escaped a unit
mesh cell at least once (see Fig.~\ref{rdpdf}b)), long range correlations persist
and the gel bead positions have not decorrelated much. Looking at even longer time
scales ($\Delta>40$), on which the tracer typically traverses several unit cells,
the long range  position correlations persist while the $G_T(0,t)$ region already
shows oscillations. A similar but even more drastic picture is found in the case
$\sigma=3$, where, in contrast, the tracer typically remains inside a single unit
cell while escaping it only occasionally.

From these results we are able to draw a full physical picture for the
emergence of transient anomalous diffusion in flexible gels. For tracer
particles of size comparable to the gel beads and for gel structures with
mesh size of the order of several times the tracer size the observed
diffusion is normal, with a a renormalized diffusion coefficient reflecting
the average effect of tracer-gel collisions. While the tracer size grows the
probability to collide with the gel increases strongly and the free diffusion
coefficient is reduced, suggesting that on the time scale of the gel breathing
modes the correlations in the tracer dynamics are not yet relaxed. The resulting
coupling creates persistent long-range correlations on time scales on which the
tracer already crosses several unit cells. Since the memory of the location of
the tracer before an escape to an adjacent cell is not destroyed, the same holds
true for the entrance to a new cell. That is, the beads of the cell into
which the tracer is about to cross already experience effects of the vicinity of
the tracer before the tracer actually enters. In turn, this demonstrates that the
intermediate-asymptotic dynamics are in fact collective, giving rise to distinct
transient subdiffusion. On fully asymptotic time scales, on which the tracer
visits a large number of mesh cells, the correlations are completely destroyed,
resulting in the terminal normal diffusion with an effective diffusion coefficient
scaling as $D_{eff}^{\infty}\simeq L_0^2/\overline{\tau}_{esc}$, where
$\overline{\tau}_{esc}$ is the mean escape time from a unit mesh cell.

\emph{Conclusion.} We used extensive Brownian dynamics simulations to study the
time dependent correlations for the motion of tracer particles in a flexible,
thermally agitated gel. Going beyond the single trajectory analysis solely of the
tracer dynamics we demonstrated that the distinct transient subdiffusion, that is
frequently observed in  experiments and simulations, is effected by the collective
fluctuations of both the tracer and the vicinal gel beads. On the scale of the
typical escape time of the tracer from a unit cell the dynamics for larger beads
turn out to still be appreciably correlated. While memory effects are a common
argument invoked in explaining anomalous diffusion phenomena, we here obtained
concrete evidence identifying its physical origin. Namely, this memory arises
from persistent many-body correlations in the entangled tracer-gel dynamics.
These correlations
for larger tracer particles clearly reach beyond the nearest neighbor gel beads.
This new physical insight obtained from the van Hove cross correlation function
and the displacement distribution sheds new light on the
elusive problem of tracer-gel interaction in the experimentally important limit
when the tracer size is comparable to the typical mesh size.
We are confident that these results will inspire new experimental and theoretical
investigations of tracer motion in structured matrices.

\emph{Acknowledgment.} AG  acknowledges funding through  an Alexander
von Humboldt  Fellowship. RM acknowledges funding from  the Academy of
Finland (FiDiPro  scheme). The  authors gratefully thank  the National
Institute of Chemistry, Slovenia for providing extensive access to the
institutes computational facilities.


\begin{thebibliography}{99}
\bibitem{comm} For $\sigma=2.5$ terminal normal diffusion is reached approximately
at $\Delta\sim200$ while in the case of $\sigma=3$ a clear crossover to linear
scaling was not observed after averaging over a trajectory of length $T=10^4$.
For $\Delta>100$ the results become noisy due to deteriorating statistics.
\end{thebibliography}
\end{document}